\documentclass[twocolumn]{aastex62}
\usepackage{amsmath}

\newcommand\ma{M_{Ax}}
\newcommand\ms{M_{s}}
\newcommand\gm{(\Gamma - 1)}
\newcommand\etag{\frac{\Gamma}{\gm}}
\newcommand\tn{\tan^2 (\theta_{Bn})}
\newcommand\cc{\cos^2(\theta_{Bn})}
\newcommand\ocii{\Omega_{ci}^{-1}}
\newcommand{\dHybridR}{{\it dHybridR}}

\newcommand{\dxlb}{\frac{\partial }{\partial x} \left[}
\newcommand{\lb}{\left[}
\newcommand{\rb}{\right]}

\newcommand{\rbud}{\right]}

\graphicspath{{./}}

\submitjournal{ApJ}

\shorttitle{Shock Heat Flux}
\shortauthors{Haggerty et al.}

\begin{document}

\title{The Importance of Heat Flux in Low Mach Number, Quasi-Parallel Collisionless Shocks}

\correspondingauthor{Colby C. Haggerty}
\email{ColbyH@hawaii.edu}

\author[0000-0002-2160-7288]{Colby C. Haggerty}
\affil{Institute for Astronomy, University of Hawaii, Manoa, 2680 Woodlawn Dr., Honolulu, HI 96822, USA}

\author[0000-0003-0939-8775]{Damiano Caprioli}
\affiliation{Department of Astronomy and Astrophysics, University of Chicago, 5640 S Ellis Ave, Chicago, IL 60637, USA}

\author[0000-0002-5938-1050]{Paul A.~Cassak}
\affiliation{Department of Physics and Astronomy and the Center for KINETIC Plasma Physics, West Virginia University, Morgantown, West Virginia 26506, USA}
\affiliation{Department of Physics and Astronomy, Clemson University, Clemson, SC, 29634, USA}

\author[0000-0001-6330-1650]{M. Hasan Barbhuiya}
\affiliation{Department of Physics and Astronomy and the Center for KINETIC Plasma Physics, West Virginia University, Morgantown, West Virginia 26506, USA}
\affiliation{Department of Physics and Astronomy, Clemson University, Clemson, SC, 29634, USA}

\author[0000-0002-4313-1970]{Lynn Wilson III}
\affiliation{NASA Goddard Space Flight Center, Heliophysics Science Division, Greenbelt, MD, USA}

\author[0000-0002-2425-7818]{Drew L. Turner}
\affiliation{Johns Hopkins University Applied Physics Laboratory: Laurel, MD, USA}

\begin{abstract}
Collisionless plasma shocks are a common feature of many space and astrophysical systems. They are sources of high-energy particles and non-thermal emission, channeling as much as 20\% of the shock's energy into non-thermal particles.
The generation and acceleration of these non-thermal particles have been previously studied and shown to affect the shock hydrodynamics to the zero-th order.
In this work, we use self-consistent hybrid particle-in-cell simulations to examine the effect of self-generated, non-thermal ion populations on the nature of collisionless, quasi-parallel shocks.
Accelerated, non-thermal particles downstream of the shock diffuse into the upstream region, taking energy away from the shock, which increases the compression ratio, slows the shock down, and flattens the non-thermal population's spectral index for lower Mach number shocks.
We show that this enhances shock compressibility when the heat flux is included in the Rankine-Hugoniot jump conditions, results that are roughly consistent with previous theories of ``cosmic-ray modified shocks''.
Additionally, the simulation data shows that heat flux and enthalpy flux cancels out in the upstream region, yielding a relatively simple, alternative closure for the jump conditions which accurately predict for the shock speed and compression ratio.
The results have the potential to explain discrepancies between predictions and observations in a wide range of systems, such as inaccuracies in predictions of arrival times of coronal mass ejections  
and the conflicting radio and x-ray observations of intracluster shocks.
These effects will likely need to be included in fluid modeling to predict shock evolution accurately. 
\end{abstract}


\section{Introduction} \label{sec:intro}
Shock waves are ubiquitous plasma phenomena occurring in many heliospheric and astrophysical contexts, including systems such as Earth's bow shock, interplanetary coronal mass ejections, the heliospheric termination shock, binary stellar wind interactions, supernovae, and galaxy cluster mergers \citep{burgess+15,lee+12,ellison+07,ryu+03,kang+12}.
There are two characteristic features that all of these systems share: 

1. They occur in environments where the collisional mean free path is much larger than the other fundamental plasma length scales, e.g., the proton gyro-radius.

2. They are sources of energetic particles and non-thermal emission.

These characteristics of astrophysical plasma shocks are potentially important to their evolution and can cause a significant deviation in their behavior from collisional, fluid shocks.

Previous simulations and theoretical works have connected these two features, showing that collisionless shocks generate and accelerate non-thermal particles (e.g., \cite{bell78a,bell04,caprioli+14a}).
The reformation of the shock initially energizes the non-thermal particles until reaching an energy $E$ for which the particle can pass back and forth across the shock front ($E/E_{sh} \gtrapprox 10$, with $E_{sh} = m_i u_1^2/2$ being the kinetic energy of the upstream bulk flow in the shock frame, where $m_i$ is the ion mass 
and $u_1$ is the upstream velocity in the shock frame) \citep{caprioli+15}.
Above this energy, non-thermal particles can freely move upstream and downstream of the shock and continue to be energized by successive collisionless scatterings between these two regions, in a process referred to as diffusive shock acceleration, or DSA \citep{krymskii77,bell78a,blandford+78}.

The non-thermal population, while few in number, can account for order 10\% of the shock's ram energy density ($n_1 E_{sh}$, where $n_1$ is ion number density in the upstream region) \citep{volk+05,parizot+06,caprioli+08,morlino+12,slane+14,johlander+21}.
The relatively large fraction of energy channeled into non-thermal particles, coupled with the collisionless nature of the shock, implies that the ion distribution functions are non-Maxwellian, which violates the frequently invoked assumption of fluid closures that assume Maxwellianity.
Because of the energetic non-thermal population, the hydrodynamic shock treatment neglecting these effects is an incomplete description of the properties of the shock \citep{drury83,morlino+12,bret20,haggerty+22}.
Previous studies have worked towards developing a more complete description and predicting the effects of the self-generated non-thermal population(usually referred to as Cosmic Rays, CRs) on the shock dynamics; these shocks are thus referred to as CR-modified shocks \citep{drury+81a,drury+81b,drury83,blandford+87,jones+91,malkov+01, kang+06,kang+07, caprioli+09a,caprioli+09b,caprioli+10a,diesing+21}.
In quasi-parallel shock configurations, these energetic, non-thermal particles are accelerated through repeated interactions with the shock front and successive intervals of shock drift acceleration \citep{caprioli+15}.
Additionally, more recent works have begun including the effects of non-thermal particles on the physics of fluid shock models, which can reproduce some features predicted by CR-modified shock theory, with results that strongly depend on the non-thermal particle energy \citep[e.g.,][]{bai+15, vanmarle+18, pfrommer+17}. 
However, this quantity must be set a posteriori.

This work considers the effect of a self-generated, non-thermal population on the nature of lower Mach number  shocks ($M_s \sim M_A \lessapprox 5$) from first  principles using self-consistent hybrid particle-in-cell simulations.
We show that the non-thermal particles downstream of the shock that diffuse into the upstream region take energy away from the shock, increasing its compressibility and slowing it down, as predicted in CR-modified shocks.
We show that upstream of the shock, these escaping supra-thermal and non-thermal particles create a large heat flux density\footnote{This heat flux is associated with self-generated non-thermal particles from the shock, rather than a pre-seeded population that the shock is passing through, like the electron strahl in the solar wind.}, which has been suggested to be important in previous works \citep{blandford+87,alves+22}. However, this large heat flux is counterbalanced by an equal and opposite enthalpy flux density in the upstream region, a condition which can be used to develop a simple and accurate closure for the Rankine-Hugoniot jump conditions.
This process increases the compression ratio and flattens the self-generated non-thermal population's spectral index for lower Mach number shocks, as was proposed for CR-modified shocks.
This result implies that collisionless shocks will travel on the order of $\approx 10\%$ slower than hydrodynamic predictions, which neglect the heat flux, for systems with efficient non-thermal particle acceleration, a seemingly universal feature of quasi-parallel shocks.
We present a simple revised theory for the Rankine-Hugoniot jump conditions, including this effect, which yields results consistent with simulations.
Finally, this effect can account for discrepancies between predictions and observations in shocks ranging from arrival times of coronal mass ejections to non-thermal spectra of Galaxy cluster mergers, attesting to the potential universality of this result.
The former has implications for space weather prediction and readiness.

\section{Non-Thermal Particles in Quasi-Parallel Shocks}
Non-thermal particles are efficiently accelerated by quasi-parallel shocks, i.e., shocks where the direction of propagation is quasi-parallel to the magnetic field in the upstream plasma, quantified by the angle $\theta_{B_n} = \cos^{-1}(|B_x|/|\mathbf{B}|) \lessapprox 45^{\circ}$, where $\mathbf{B}$ is the magnetic field vector and the $x$-direction is in the direction of shock propagation and normal to the shock front (e.g., \cite{morlino+12,caprioli+14a,johlander+21}.
In these systems, non-thermal particles with $E \gtrsim 10 E_{sh}$ are free to travel upstream and downstream relatively unaffected by the shock front \citep{caprioli+14a,caprioli+15,caprioli+18}.
While the non-thermal particles downstream of the shock are roughly isotropic, the upstream non-thermal population diffuses into the upstream, which exerts pressure on the inflowing upstream plasma, weakening the shock in two ways \citep{ellison+07,caprioli+08,haggerty+20,caprioli+20}: 
reducing the inflow speed and increasing the upstream temperature,
which combine to reduce the sonic and magnetosonic Mach number \citep{burgess+15}.
\begin{figure}[ht]
\plotone{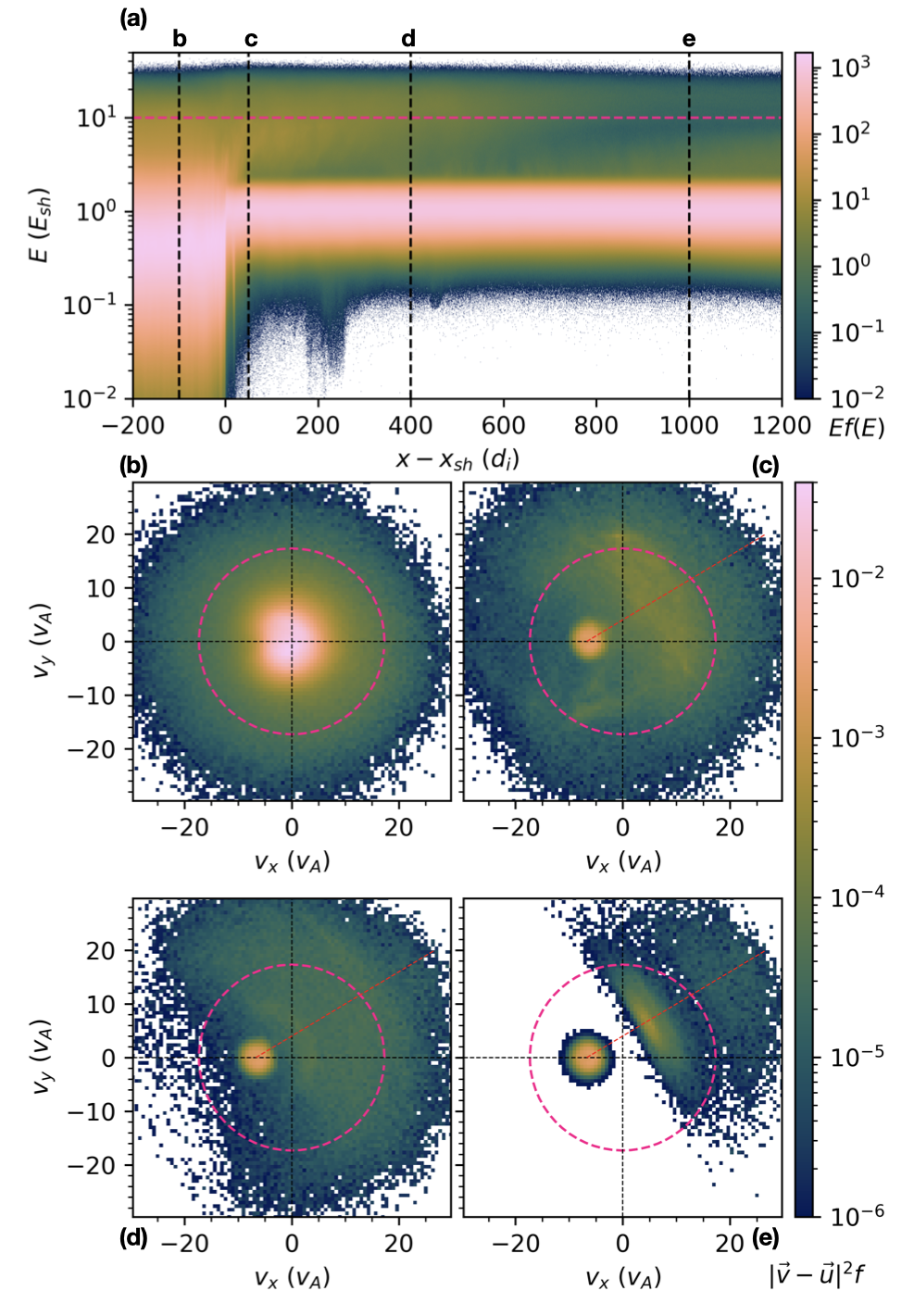}
\caption{
(a) The reduced ion energy distribution for a $2 d_i$ slice in the $y$ direction as a function of shock normal direction and energy.
(b) - (d) Ion distribution functions at the locations shown by the dashed lines in (a), scaled by $|\mathbf{v} - \mathbf{u}|^2$. The distribution functions are plotted in the $v_x$, $v_y$ plane and integrated over the $v_z$ direction.
All the distribution functions are averaged in time over $50 \ocii$ at $0.5 \ocii$ intervals and are determined in the shock frame.
}
\label{fig:M}
\end{figure}
Of the two effects, the increased upstream temperature is more significant because of the non-thermal particles' high energy and its $v^2$ dependence in the second moment of the distribution.
Simulations and observations have consistently shown that in the downstream and in near-upstream\footnote{Here ``near upstream" means within several diffusion lengths of the non-thermal ions containing most of the energy in the non-thermal distribution.} regions of the shock, on the order of 10\% of the shock's available energy density is channeled into non-thermal particle energy density \citep{caprioli+14a,caprioli+14b,caprioli+14c,haggerty+20,morlino+12,ackermann+13}, $\xi \equiv P_{nt}/\rho_{1}u_{1}^2 \approx 0.1$ where $P_{nt}$ is the pressure in the non-thermal particles\footnote{We use the subscripts `1' and `2' for quantities measured upstream and downstream, respectively.} and $\rho_1$ is the upstream mass density. 

The increased upstream pressure from the non-thermal distribution is evident in Figure~\ref{fig:M}, which shows results from a quasi-parallel shock simulation performed with the  hybrid particle-in-cell {\it dHybridR} code (see Appendix~\ref{apdx:sims} for details on the code and the simulation used in this work).
The extensive non-thermal population can be seen in the reduced distribution functions plotted in Figure~\ref{fig:M}a with energy vs shock normal and b-d, where $f(v_x,v_y)$ has been integrated over $v_z$ and is weighted by the difference in velocity to the upstream bulk flow squared $|\mathbf{v} - \mathbf{u}|^2$. 
Each of the sub-panels corresponds to the different locations as indicated by the vertical dashed lines in panel (a). 
All of the distribution functions in Figure~\ref{fig:M} are averaged in time (over 100 consecutive cuts in time spaced out over 50 $\Omega_{ci}^{-1}$) and space (a $2d_i^2$ box at $y=12.5d_i$).

The red dashed line indicates the direction of the upstream magnetic field.
The pink dashed line circle corresponds to $\frac{1}{2}m_i(v_{x}^2 + v_y^2) > 20E_{sh}/3$, showing the approximate boundary between the ions determined to be ``thermal'' (inside the circle) and ``non-thermal'' (outside the circle).
We note that it is the approximate boundary of $E > 10 E_{sh}$ because the distribution functions in b - e are integrated over the $v_z$ component, which accounts for the $2/3$ factor.
In both Figure~\ref{fig:M}~(b) and (c) we see a distinct population of energetic ions with $E > 10 E_{sh}$ which is roughly unchanged between the downstream and near upstream.
These energetic ions close to the shock front are accompanied by a significant fraction of lower energy particles, i.e., the ion foreshock region, which are attenuated with increasing distance from the shock, where only a limited, field-aligned fraction can reach.
This intermediate-energy population of reflected particles are not part of the core inflowing ion distribution; however, for this work, we will assume that they are part of the ``thermal'' population, an assumption that will be examined in future works.
\begin{figure}[ht]
\plotone{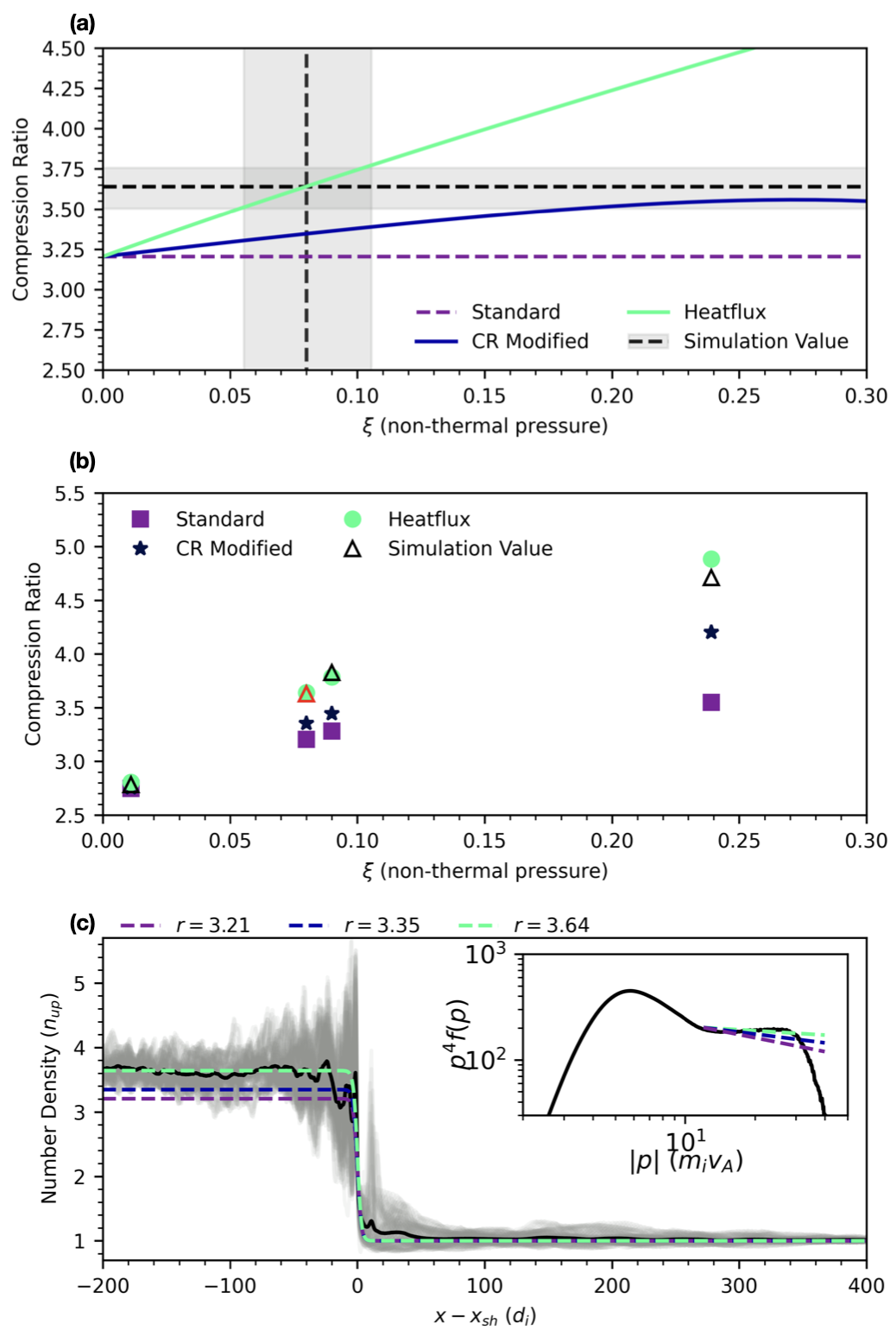}
\caption{(a): predictions for the shock compression ratio as a function of the normalized non-thermal pressure ($\xi$) from different theories as given in the legend. The black dashed lines show the simulation measured non-thermal pressure (vertical) and compression ratio (horizontal), and the theory presented in this work in teal.
(b): a scatter plot for the different compression ratio predictions for 4 different shock simulations as a function of measured non-thermal pressure ($\xi$) and Mach number. The red triangle denotes the simulation presented in the other figures.
(c): the ion number density plotted as a function of shock normal distance (time averaged as in Figure~\ref{fig:M}~\&~\ref{fig:avg_jumps}). 
The insert shows the average downstream distribution function (multiplied by $p^4$) as a function of momentum.
The predicted non-thermal power law indices are shown by the dashed lines based on standard DSA theory from the corresponding compression ratios.}
\label{fig:compress}
\end{figure}

Previous works have considered the dynamical role of these non-thermal particles, commonly referred to as CRs, in the astro literature \citep[e.g.,][]{drury+81a, drury+81b, drury83,blandford+87, jones+91, ellison+00, malkov+01, blasi02, amato+05, amato+06, kang+02, kang+05, caprioli+09a,caprioli+09b,caprioli+10a}.
The primary approach to these works is to split the particles into thermal and non-thermal contributions. The non-thermal population is assumed to be decoupled from the shock but to produce a precursor region where the inflowing/upstream plasma is preheated and compressed before reaching the discontinuity (often referred to as the sub-shock). 
The thermal plasma should follow the MHD RH jump conditions at the sub-shock, as the non-thermal particles do not see the discontinuity (i.e., the CR distribution function and hence its moments are continuous).
While the thermal plasma just upstream of the shock will be compressed and heated (resulting in a reduced sub-shock Mach number), so long as the shock is sufficiently strong, the total shock compression (between the downstream and far upstream) will be larger than the standard fluid prediction (and even become larger than 4). 
The dynamics of the non-thermal particles are treated with the Parker  transport equation including advection and diffusion \citep[e.g.,][]{skilling75a,bell78a,blandford+78}, and the two sets of equations are used to solve the compression ratio as a function of the non-thermal pressure.

The hybrid simulations are found to be qualitatively consistent with the predictions of CR-modified shocks.
Panel (c) of Figure~\ref{fig:compress} shows a time averaged cut of the ion number density as a function of distance from the shock front; from this the compression ratio of the shock can be inferred as the asymptotic upstream density is 1.
The compression ratio in the simulation reaches an average value of about $r \sim 3.64$, much larger than the prediction based on the asymptotic Rankine-Hugoniot jump conditions ($r \sim 3.21$, shown by the purple dashed line in Figure~\ref{fig:compress}).
However, the simulation compression ratio is also larger than the CR-modified shock prediction, using a non-thermal ion pressure of $\xi = 0.08$, as inferred from the simulation and shown in Figure~\ref{fig:Pnt}; the CR-modified shock prediction is shown as the blue dashed line in Figure~\ref{fig:compress}.
Panel (a) of Figure~\ref{fig:compress} underscores this point showing the predicted compression ratio as a function of non-thermal pressure for the RH-MHD (dashed purple) and the CR-modified (solid blue) predictions. 
The average compression ratio and non-thermal pressure from the simulation are shown by the intersecting dashed black lines with grey shaded lines corresponded to the standard deviation determined over a region $250 d_i$ around the shock front averaged over 20 time slices spread out over $100 \Omega_{ci}^{-1}$.

Additionally, the spectral index of the non-thermal, power law distribution appears roughly consistent with this enhanced compression, i.e., the power-law distribution following $f \sim p^{-q}$ where $q$ is determined solely by the compression ratio $q = 3r/(r-1)$ \citep{krymskii77,bell78a}. 
This is shown in the inset figure, where the dashed lines correspond to standard DSA predictions
\footnote{Note that we are omitting the hydrodynamic and spectral index corrections discussed in \cite{haggerty+20} and \cite{caprioli+20}, because of the modest magnetic field amplification associated with lower Mach number shocks.}.
However, it should be noted that the extent of the power-law tail will only reach an order of magnitude in energy space, and that larger simulations should be performed for a more robust verification\footnote{The function in the insert shows the average downstream distribution (multiplied by $p^4$) as a function of momentum, plotted in this way to highlight the power law slope.}

This simulation shows that the non-thermal pressure is increasing the compression ratio, to an even greater extent than was predicted by CR-modified shocks.
In the following section we present an alternative closure to the RH-jump conditions which is in agreement with the compression ration found in the simulation.

\section{Heat Flux in the Jump Conditions}
An alternative approach to predicting the compression ratio can be developed by reexamining the RH jump conditions from a kinetic perspective.
Much of the analysis of collisionless shocks is derived from the single fluid, MHD limit.
In this limit, the jump condition describing the energy flux density on either side of the shock omits the heat flux density vector, a third-order moment of the distribution function, defined for an arbitrary distribution as $Q_j = (m/2)\int (v_j - u_j)|\mathbf{v} - \mathbf{u}|^2f d^3v$ for a species with mass $m$.
However, this limit is not appropriate for the upstream region of a collisionless shock when non-thermal particles are present, as is demonstrated by the distribution functions shown in Figure~\ref{fig:M}(a)-(d).
Close to the shock front, the significant presence of energetic, non-thermal particles suggests that heat flux density is comparable to the flux of both the bulk flow energy density and thermal energy density, or more precisely, the enthalpy flux density (see Appendix~\ref{apdx:QPflux} for details).

\begin{figure}[t]
\includegraphics[width=0.45\textwidth,trim={10 0 10 0},clip]{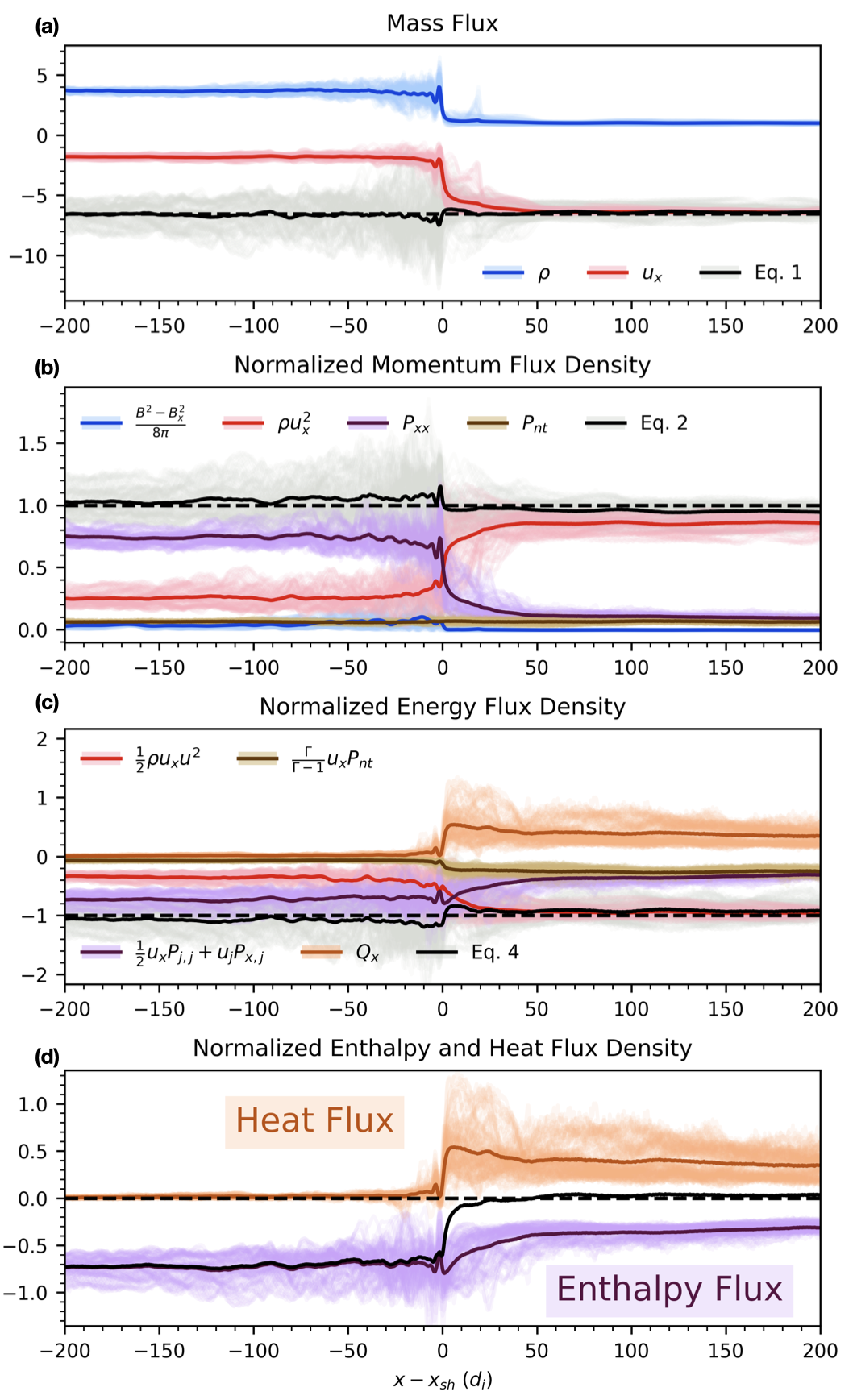}
\caption{
    Time-averaged Rankine-Hugoniot jump conditions. 1D cuts normal to the shock and centered around the shock front time-averaged as discussed in Figure~\ref{fig:M}. The black lines in each of the panels correspond to the sum of the fluxes in Eq.~\ref{eq:mass}, \ref{eq:momentum}, and \ref{eq:energy} for panel (a), (b), and (c) respectively. The colored lines correspond to the different terms in each of the equations and the brown lines show the non-thermal contribution to the pressure and enthalpy flux density separately. Panel (d) shows the sum (black line) of the heat flux density (orange) and the total enthalpy flux density (purple). The two almost perfectly counterbalance each other in the upstream region, showing that the upstream, non-thermal particles do not contribute to the jump conditions.
}
\label{fig:avg_jumps}
\end{figure}

The importance of the heat flux on the jump conditions is quantified directly in Eqs.~\ref{eq:mass}-\ref{eq:energy} and in Figure~\ref{fig:avg_jumps}, which shows the various quantities determined from the simulations, including those that make up the jump conditions and the heat flux.
Each quantity is determined in the shock rest frame (as measured in the simulation), and the figure employs the same time averaging used in Figure~\ref{fig:M}.
Figure~\ref{fig:avg_jumps} shows the individual components and total of the following jump conditions:
\begin{align}
\left [ \rho u_x \right ] = 0, \label{eq:mass}\\
\left [ \rho u_x^2 + P_{xx} + \frac{B_y^2}{8\pi}\right ] = 0,\label{eq:momentum}\\
\left [ \rho u_x u_y - \frac{B_xB_y}{4\pi}\right ] = 0,\label{eq:momentumy}\\
\left [  
\frac{1}{2}u_x\rho u^2 + \frac{1}{2}u_xP_{j,j} + u_{j}P_{j,x} + Q_x
\right  ] = 0,\label{eq:energy}
\end{align}
With Eqs.~\ref{eq:mass}, \ref{eq:momentum} \& \ref{eq:energy} corresponding to panels (a), (b), and (c) of Figure~\ref{fig:avg_jumps}, respectively, and where the Einstein summation convention is used for repeated dummy indices. 
The square brackets correspond to the difference between any point in the upstream and downstream regions where the assumption of stationary remains valid.
The equations are written in the de~Hoffmann-Teller frame so that all flows are parallel to the magnetic field direction (i.e., $\mathbf{u}\times \mathbf{B} = 0$) \citep{DeHoffmann+50}; additionally, the \textit{x} component of the magnetic field must be constant through the shock.
In Eqs.~\ref{eq:mass} -- \ref{eq:energy},
$P_{j,k}$ is the $j,k$th element of the total pressure tensor of the ions (both thermal and non-thermal) and electrons\footnote{Note for the simulations: the ion pressure is determined by directly measuring the second moment of the distribution function, while the electron pressure is set by the adiabatic electron equation of state, with $n_i = n_e$.}.
$Q_x$ is the heat flux density of the full ion distribution normal to the shock (electrons cannot contribute to the heat flux in the hybrid model because they are assumed to be isotropic). 
Note that these equations converge to the typical MHD jump conditions when the pressure tensor becomes isotropic and the heat flux becomes negligible ($Q_x \sim 0$).

The momentum and energy flux in Figure~\ref{fig:avg_jumps} confirm what was evident from the upstream distribution functions of Figure~\ref{fig:M}; the non-thermal particles have a zeroth order contribution to the upstream terms in the jump conditions, in the enthalpy flux density (purple line, panel (c)) and heat flux density (orange line, panel (c)).

While each of these terms are comparable to either the momentum or bulk energy flux density in the upstream flow, the energy flux panel reveals a critical relationship between these terms.
In the upstream region, the heat flux density is nearly equal in magnitude and opposite in sign to the enthalpy flux density that has been enhanced by the non-thermal particles.
Physically, this corresponds to the non-thermal particles traveling away from the shock and carrying energy with them. 
This claim is supported by the brown line in panel (c) of Figure~\ref{fig:avg_jumps} which shows the normalized approximate enthalpy flux density due to the non-thermal population (as detailed in Appendix~\ref{apdx:QPflux} and \ref{apdx:jumpheatbal} ), which becomes comparable to the total enthalpy flux density (purple line). 
In the upstream region, the sum of all of the components (black line) is almost exactly equal to the ram energy density flux (red line), which confirms that the heat and enthalpy flux densities negate one another.
This cancellation can be interpreted as the balancing of the non-thermal diffusive flux away from the shock with the advection of the non-thermal population towards the shock with the thermal population's bulk flow.
This result suggests that the non-thermal population does not contribute to the upstream energy budget.
However, this cancellation allows us to close the RH-jump conditions without invoking the precursor or describing the non-thermal particles using the Parker equation as done in previous predictions for CR-modified shocks \citep{caprioli+09a}.

\section{Including Heat Flux into the shock hydrodynamic predictions}
Note that the total mass density and bulk flow are not appreciably altered by the non-thermal particles, as can be seen in 
panel (a) of Figure~\ref{fig:avg_jumps}. 
The enthalpy/heat flux cancellation will modify the jump conditions in two key ways.
First, the pressure in Eq.~\ref{eq:momentum} (the momentum flux equation) is only the thermal pressure $P_g$.
This is because the pressure of the non-thermal population is the same in the upstream and downstream very close to the shock, as they are free to pass the shock with impunity (as demonstrated by Figure~\ref{fig:Pnt} in the Appendix~\ref{apdx:QPflux}).
Second, the heat flux density is removed from Eq.~\ref{eq:energy} along with the non-thermal enthalpy flux density in the upstream region.
This leaves a non-thermal contribution to the enthalpy flux density in the downstream side of the equation, which acts as an energy sink. 
Combining these conditions yields the following set of equations:
\begin{equation}
\rho_1 u_{x,1} = \rho_2 u_{x,2}\label{eq:mflux}
\end{equation}
\begin{equation}
\rho_1 u_{x,1}^2 + P_{g,1} + \frac{B_{y,1}^2}{8\pi} = 
\rho_2 u_{x,2}^2 + P_{g,2} + \frac{B_{y,2}^2}{8\pi},\label{eq:pflux}
\end{equation}
\begin{equation}
\rho_1 u_{x,1}u_{y,1} - \frac{B_xB_{y,1}}{4\pi} = 
\rho_2 u_{x,2}u_{y,2} - \frac{B_xB_{y,2}}{4\pi},\label{eq:bflux}
\end{equation}
\begin{equation}
\begin{aligned}
u_{x,1} \left ( \frac{1}{2}\rho_1 u_1^2 + \frac{\Gamma}{\Gamma - 1}P_{g,1} \right ) = \\
u_{x,2} \left ( \frac{1}{2}\rho_2 u_2^2 + \frac{\Gamma}{\Gamma - 1}(P_{g,2} + P_{nt,2}) \right ).\label{eq:Eflux}
\end{aligned}
\end{equation}
Here the ``1'' corresponds to the roughly constant values upstream beyond roughly $50 d_i$, and ``2'' to the downstream region.
$\Gamma = 5/3$ is the non-relativistic, adiabatic index, which holds for any isotropic distribution function, as motivated by Figure~\ref{fig:M}~(b) and (c).
Note that the non-thermal particles contribute their pressure to the energy equation in the downstream region but not in the upstream region, or in any other equation.
Physically, this increases the compressibility of the shock, causing the shock to slow down.
The derivation starting form the Vlasov equation, splitting the ``thermal'' and ``non-thermal'' terms leading to Eq.~\ref{eq:Eflux} is provided in detail in Appendix~\ref{apdx:jumpheatbal}.

We can explicitly solve for the compression ratio using the modified jump conditions with the inclusion of the normalized non-thermal pressure $\xi$, and the exact solution is derived in Appendix~\ref{apdx:jumpheatbal} (Eq.~\ref{eq:r:apdx}).
However, the prediction can be simplified in the limit of realistic non-thermal efficiencies ($\xi \sim 0.1$) for parallel shocks:
\begin{equation}
    r \approx
    \frac{(\Gamma + 1)M^2}{2 + (\Gamma - 1)M^2} + 
    \frac{\Gamma (\Gamma + 1)}{\Gamma - 1}\xi\label{eq:r}
\end{equation}
The first term in Eq.~\ref{eq:r} is the standard MHD RH prediction for an unmodified shock, and the second term is positive and linear in $\xi$, which means that the non-thermal population will increase the comparability of the shock (consistent with Figure~\ref{fig:compress}).
As $\xi \rightarrow 0$, Eq.~\ref{eq:r:apdx} (and Eq.~\ref{eq:r}) converges to the compression ratio prediction from the Rankine-Hugoniot jump conditions in the MHD limit.
Finally, Eq.~\ref{eq:r} can be further simplified in the limit of large $M_{A}$ and $M_s$:
\begin{equation}
    r \approx 4 (1 + \Gamma \xi), \label{eq:strong_shock_limit}.
\end{equation}
It should be noted that this accounts for only the hydrodynamic effects from the heat flux but omits the contribution of the non-thermal postcursor discussed in \cite{haggerty+20}, which will be important for higher Mach number shocks.

In observations and simulations of quasi-parallel collisionless shocks, $\xi$ is consistently found to be on the order of $0.1$ \citep{volk+05,parizot+06,caprioli+08,caprioli+14a,haggerty+19a,johlander+21}. Applying this to Eq.~\ref{eq:strong_shock_limit}, we find that the compression ratio of strong quasi-parallel shocks based on these updated Rankine-Hugoniot predictions will increase the compression ratio by approximately $\Gamma/10 = 1/6 \approx 17\%$.
Using Eq.~\ref{eq:r:apdx} and the simulation measured value of $\xi \approx .08$, we find striking agreement between the measured and predicted compression ratio in the simulation (teal line in panels (a) and (c) of Figure~\ref{fig:compress}).
We extend this analysis and prediction to 3 additional parallel shock simulations with varying Mach numbers (detailed in Appendix~\ref{apdx:sims}). For each simulation, we determine the late-time compression ratio and the normalized non-thermal pressure $\xi$, and we calculate the Mach number. Then the compression ratio is calculated using the standard approach (purple squares), CR modified prediction (blue stars), the heat flux prediction (teal circles), and compared to the simulation measured
value (black triangles) in panel (b) of Figure~\ref{fig:compress}. The location of each point on the horizontal axis roughly scales with the Mach number squared, and the red triangle denotes the canonical simulation in this work. We find good agreement for the range of initial conditions.

Considering this analysis, a natural question arises: Why can the jump conditions not be calculated using the values sufficiently far upstream of the shock, such that the non-thermal particles would not have yet reached (which would be equivalent to the prediction of the purple dashed line in Figure~\ref{fig:compress})?
Applying the jump conditions at this location yields inaccurate results likely because of the time-dependent flux of non-thermal particles in the far upstream (ions at $> 500 d_i$ in Figure~\ref{fig:M}a).
The non-thermal pressure is continuous across the shock with relatively small variations in space and time within the relative vicinity of the shock front (see Figure~\ref{fig:Pnt} in Appendix~\ref{apdx:QPflux}), while far upstream, non-thermal particles are escaping farther away from the shock.
This escaping flux should violate the stationary assumption and invalidates one of the assumptions of the jump conditions in the far upstream region.
The escaping ions carry a fraction of the shock energy but do not appreciably pre-compress or heat the inflowing plasma.
This time-dependent contribution is omitted from CR-modified shock theory (though see \citet{drury+81a, drury+81b, zirakashvili+08, caprioli+09b}), typically on the basis of being negligible, however this contribution has a larger impact on the hydrodynamics for lower Mach number shocks, such as those presented in this work.

Finally, there remains the issue of the discrepancy in the predictions for the compression ratio between CR-modified work (blue lines in Figure~\ref{fig:compress}) and the heat flux work presented in this manuscript (mint lines in Figure~\ref{fig:compress}). 
The cause for this discrepancy is potentially due to the assumptions of CR-modified shock theory applied to the relatively low Mach number shock presented in this work.
CR-modified shock theory treats the non-thermal particles using a separate transport equation coupled to the shock's upstream and downstream regions \cite{skilling75a,amato+05,caprioli+09a}.
This formalism assumes particles scatter off magnetic fluctuations and tend towards isotropy over length scales comparable to the non-thermal particles gyro-radius.
The transport equation is closed by ordering in terms of the anisotropy of the distribution function and then neglecting terms higher than the first order.
This closure requires relatively strong magnetic fluctuations; however, lower Mach number shocks produce smaller amplitude magnetic fluctuations \citep{caprioli+14b}.
Because of this, higher-order anisotropic terms in the distribution function may not be negligible, potentially resulting in the discrepancy shown in Figure~\ref{fig:compress}.
However, as the Mach number increases along with the corresponding magnetic fluctuations, we anticipate that the CR-modified shock predictions will better agree with simulations as demonstrated in \citet{Haggerty+18}.

In contrast, the heat flux/enthalpy flux closure used in this work is derived based on empirical evidence from the simulation and is exact in that higher-order moments do not need to be neglected to close the equations.
The closure presented in this work is comparatively more straightforward than the CR-modified counterpart, as the Parker transport equation was not needed to close the set of equations.
While this theory is likely useful for accurately predicting shock hydrodynamics, it does not consider the form of the non-thermal particle distribution function, which could still be determined using the CR-modified shock theory.

\section{Implications for Space and Astrophysical Shocks}
The hydrodynamic modifications outlined in this paper are expected to be a generic feature of any collisionless shock with a non-thermal population (e.g., quasi-parallel configurations), which implies that these results are potentially impactful on disparate systems including solar, heliospheric, and astrophysical environments.
The non-thermal enhancement of the shock compressibility has three immediate implications:
First, the compression ratio of quasi-parallel shocks is larger than the standard hydrodynamic prediction, reaching or surpassing the hydrodynamic maximum of 4, even for low Mach number shocks.
Second, quasi-parallel shocks travel slower than fluid theory predicts, where the fractional decrease in speed is comparable to the fraction of energy channeled into non-thermal particles (i.e., on the order of 10\%, as the speed of the shock in the unshocked frame is given by $u_{sh} \propto 1/(1 - 1/r)$).
Third, the slopes of non-thermal, power law distributions are flatter for a given Mach number, as the slope increases with the compression ratio as predicted by 
DSA\footnote{This effect is most evident for lower Mach number shocks, in which the self-generated, amplified magnetic field is not large enough to steepen the non-thermal spectra, as is discussed in \cite{caprioli+20}. While less obvious, this effect is expected to be present in higher Mach number shocks and can likely account for the discrepancy between the predicted and measured compression ratio in \cite{haggerty+20}.}.

The compressional enhancement and associated shock speed reduction have significant implications for numerous astrophysical systems, including the lifetime and evolution of supernova remnants, stellar termination shocks, planetary bow shocks and astrophysical jets.
More locally, a potential application for this result is for the propagation of interplanetary coronal mass ejections (ICMEs), the accurate modeling of which is crucial for space weather forecasting.
Various analytical hydrodynamic and MHD models for ICMEs consistently underestimate Earth arrival times by approximately 10 hours (e.g., \citet{Dumbovic+18}).
Note that this is in contrast to empirical, data driven models frequently used for forecasting space weather which have more accurate prediction times \citep{kay+24}.
Given that ICMEs are efficient sources of solar energetic particles \citep{reames+13,webb+12,kamijima+20}, the physics described in this manuscript is expected to be applicable, and to likely account for the order 10\% decrease in ICME speeds and an average 10 hour increase in transit time (typically on the order of 100 hours) \citep{gopalswamy+01,Dumbovic+18}.

Additionally, the flattening of the power law spectrum has implications for the acceleration of CRs and high energy emission from shocks.
This effect could explain an observational discrepancy in the shocks from galaxy cluster mergers, in which the shock Mach numbers are inferred from x-ray observations, while the spectral index is inferred from radio emission (e.g., \cite{vanWeeren+19,ha+23}).
The observed spectral indices consistently correspond to a larger Mach number than is inferred from the X-ray observations.
The effect described in this manuscript explains such measurements and Eq.~\ref{eq:r:apdx} can be used to bridge these observations.
For example, for the Sausage relic (CIZA J$2242.8+5301$), the Mach number is estimated using the radio spectral index of $4.2$, which would correspond to a compression ratio of $3.6$ and a Mach number of $4.6$ \citep{vanweeren+10}, while the value estimated through temperature measurements through X-ray observations is smaller with $M \approx 2.5-3.1$ \citep{ogrean+14,akamatsu+15}.
Using Eq.~\ref{eq:r:apdx} with a Mach number of $2.7$, a compression ratio of $3.6$ would be predicted with a non-thermal/CR efficiency of only $\xi \approx 0.15$, which would completely account for the observational discrepancies.

The results presented in this manuscript may likely be impactful for other systems including planetary bow shocks, solar/stellar wind termination shocks and supernova remnants.
One potential system in which these results could be constrained and applied is Earth's day-side magnetosphere, which exhibits strong dawn-dusk asymmetries of the shape of Earth's bow shock\citep{walsh+14,dimmock+17}.
One proposed source of this dawn-dusk asymmetry is the corresponding quasi-parallel-to-quasi-perpendicular transition of the shock.
The asymmetry has been correlated with the presence of energetic ions \citep{anagnostopoulos+05,johlander+21}, which suggests that this paper's results may, in-part, account for the observed symmetry breaking.

While these results may be applicable to several different astrophysical shock systems, caution should be taken when applying them, given the complex, non-linear nature of shocks. 
In situ observations of Earth's bow shock consistently reveal the presence of shock-related transient phenomena which have zeroth-order effects on the magnetospheric system \citep{zhang+22}, such as foreshock bubbles and hot flow anomalies which can also directly affect non-thermal ion acceleration \citep{tzliu+15,tzliu+16,tzliu+18}. 
Further complications arise when including the larger magnetospheric system itself, as demonstrated by both MMS and THEMIS observations of the magnetosheath in \cite{tzliu+24}, which used a statistical set of crossings to show that a net heat flux density directed towards the upstream was consistently observed downstream of Earth's bow shock. 
While this is not found in our simulations, \cite{tzliu+24} attributed the net heat flux to a mirror force exerted by the increasing field strength towards the magnetopause, which would not be present in our simulation setup.
Including an extra constant downstream heat-flux density term in our model predicts a similar effect on the shock hydrodynamics: a percentage increase in the compression ratio, comparable to the normalized heat-flux density.
Converting our simulation heat flux density into units more applicable to Earth's bow shock (i.e., using $B\sim 5$nT, $n\sim 5/{\rm cm}^3$ and $V_{sw} \sim 400\, {\rm km/s}$) we find a heat flux density of $\sim 10^{11} \ {\rm eV}/({\rm s\, cm}^2)$, which is comparable to what was found in the observations.
This suggests that the magnetospheric boundary effect could be comparable to the results in this work.
However, a detailed accounting of the total enthalpy flux would be required to make any strong claims.
All of this supports the idea that the physics of collisionless shocks needs to be examined carefully from a fundamentally kinetic perspective to understand and model these systems accurately.
While there has been an estimate of the energy budget of the quasi-perpendicular bow shock \citep{schwartz+22}, a detailed local accounting of the quasi-parallel bow shock is limited by the available in-situ satellites in operation.
Constraining these results, as well as the kinetic dynamics of shocks more generally, will require high cadence observations of both ion and electron distribution functions; this will likely require a multi-spacecraft mission designed for the plasma environment for both the upstream  and downstream of Earth's bow shock, such as the Multi-point Assessment of the Kinematics of Shocks (MAKOS) mission \citep{goodrich+22p}.

\section{Conclusion}
In this manuscript, we demonstrate the importance of self-generated non-thermal particles in shock hydrodynamics and show that it is a ubiquitous feature of collisionless quasi-parallel shocks. We develop an alternative closure to the Rankine-Hugoniot jump conditions which include the heat flux density associated with non-thermal particles and show they agree well with simulations. We show that the generation of non-thermal particles increases a shock's compressibility, decreases the propagation speed, and flattens the associated power-law spectrum. 
We argue that this should be a generic feature of collisionless shocks with non-thermal particles and find that this effect can likely answer several outstanding issues with heliospheric and astrophysical shocks.

\section{Acknowledgments}
The Authors thanks the anonymous referee for their careful reading and comments that helped improving the manuscript.  
Simulations were performed on computational resources provided by the University of Chicago Research Computing Center, on TACC’s Stampede2 and Purdue's ANVIL through ACCESS (formally XSEDE) allocation TG-AST180008.
Some of the work was supported by the Geospace Environment Modeling Focus Group ”Particle Heating and Thermalization in Collisionless Shocks in the Magnetospheric multiscale mission (MMS) Era” led by L.B. Wilson III.
C.C.H.~was partially supported by NSF FDSS grant AGS-1936393 as well as NASA grants 80NSSC20K1273 and 80NSSC23K0099;
D.C. by NASA grants 80NSSC20K1273, 80NSSC23K1481, and 80NSSC24K0173 and NSF grants AST-2009326, PHY-2010240, and AST-2510951;
P.A.C. and M.H.B. by NASA grants 80NSSC20K1273, 80NSSC19M0146, 80NSSC23K0409 and 80NSSC22K0323, NSF grants PHY-1804428 and PHY-2308669, and DOE grant DE-SC0020294.
Finally, this work was also supported by the
International Space Science Institute (ISSI) in Bern through
ISSI International Team project No. 23-575, “Collisionless
Shock as a Self-Regulatory System” and through the Visiting
Scientist program.

\appendix
\section{Simulations and dHybridR}\label{apdx:sims}
To study the effects of heat flux and non-thermal enthalpy flux on the hydrodynamics of shocks, we perform self-consistent simulations using \dHybridR{}, a relativistic hybrid code with kinetic ions and massless, charge-neutralizing fluid electrons \citep{haggerty+19a}.
\dHybridR{} is the generalization of the non-relativistic code \emph{dHybrid} \citep{gargate+07}, which was widely used for simulating collisionless shocks \citep{gargate+12,caprioli+14a,caprioli+14b,caprioli+14c,caprioli+15, caprioli+17, caprioli+18, haggerty+19p,caprioli+19p}.

All physical quantities are normalized to their far upstream initial values, namely: mass density to $\rho_0 \equiv m_i n_0$ (with $m_i$ the ion, namely proton, mass), magnetic fields to $B_0$, lengths to the ion inertial length $d_i = c/\omega_{pi}$ (with $c$ the speed of light and $\omega_{pi}$ the ion plasma frequency),
time to the inverse ion cyclotron frequency $\ocii$,
and velocity to the Alfv\'en speed $V_{A0} = B_0/\sqrt{4\pi \rho_0}$.
The ion plasma beta is chosen to be $\beta_i = 2$, which corresponds to a thermal gyroradius of $1\, d_i$.
The system is 2.5D; 2D in real space (in the $x-y$ plane), but with all three components of momentum and electromagnetic fields retained.
The electrons are chosen to have an adiabatic equation of state, i.e., the electron pressure is $P_e\propto\rho^{5/3}$.

The simulation is initialized with a uniform magnetic field ${\mathbf{B}_0}=B_0 (\cos(\theta_{B_n})\hat{x} + \sin(\theta_{B_n})\hat{y})$ with $\theta_{B_n} = 30^\circ$ and an inflowing thermal ion population with a bulk flow $\mathbf{u}_x = - 5 V_{A0} \mathbf{x}$ in the simulation frame. 
The simulation is periodic in the $y$ direction, the right boundary is open and continuously injecting thermal particles, and the left boundary is a reflecting wall;
after tens of cyclotron times, the ions initially closest to the wall reflect and form a shock that travels in the $+\mathbf{x}$ direction, with the downstream plasma at rest in the simulation reference frame.
Note that the speed of the shock in the downstream (simulation) frame is not the same speed that goes into the determination of the Mach number.
The Mach number that enters the stationary jump conditions is measured in the shock frame which is determined empirically from the simulation ($\approx 1.7 V_A$ in the simulation frame), with the sonic and fast magneto-sonic Mach numbers defined as $M_{\rm sonic} = u_1/v_{s,1}$ and $ M_{\rm mag'sonic} = u_1/v_{m,1}$, where 
\begin{align}
    v_{s,1} &= \sqrt{\frac{\Gamma P_1}{\rho_1}}\\
    v_{A,1} &= \frac{B_1}{\sqrt{4\pi \rho_1}}\\
    v_{m,1} &=  \left [ \frac{1}{2} \left ( 
    v_{A,1}^2 + v_{s,1}^2 + \sqrt{(v_{A,1}^2 + v_{s,1}^2)^2 - 4 v_{A,1}^2v_{s,1}^2\cos^2\theta_{Bn}}
    \right ) \right]^{1/2}
\end{align}

The simulation domain size is $[L_x,\, L_y] = [2000, 25] d_i$, wide enough to account for 2D effects and long enough so that the simulation could be run $100$s of $\ocii$ without energetic particles escaping the domain.
The simulation has two grid cells per $d_i$, and each grid cell is initialized with 10000 particles per grid.
The speed of light is set to be larger than the Alfv\'en and thermal speeds ($c/V_{A0} = c/v_{thi} = 50$), as discussed in \cite{haggerty+19a}; the time step is set as $ c\Delta t = d_i/2$. 

Finally, the three other simulations used to produce the scatter plot in panel (b) of Figure~\ref{fig:compress} were run with a reduced number of 1024 particles per grid for computational feasibility. The three simulations were performed for parallel shocks with upstream flow speeds of $3,\ 5,\ {\rm and}\ 7 V_A$ as measured in the downstream frame. The compression ratio, Mach number, and non-thermal pressure were determined using the same procedure described in the manuscript and were used to predict the heat flux modified for the compression ratio.

\section{Energy Jump Conditions Including Heat Flux and Enthalpy Flux}\label{apdx:QPflux}
In this section we derive the third jump condition including kinetic contributions from the nonthermal populations (Eq.~\ref{eq:energy}).
This equation is derived from the sum of the ion and electron Vlasov equation assuming a constant solution in time:
\begin{align}
\sum_s^{i,e} \left[
\frac{\partial}{\partial x} (v_x f_s) + \bm{\nabla}_v \cdot \frac{q_s}{m_s}(\mathbf{E} + \frac{\mathbf{v}}{c}\times \mathbf{B})f_s \right] = 0.
\end{align}
By performing the calculation in the de Hoffmann-Teller frame, the magnetic force term drops out of the equation. Multiplying by $\frac{1}{2}m_sv^2$ and integrating over all velocity yields the jump condition:
\begin{align}
\sum_s^{i,e}
\frac{\partial}{\partial x} \int \frac{m_s}{2} v^2 v_x f_s d^3v = 0.
\end{align}
We derive Eq.~\ref{eq:energy} using the typical approach by letting $\mathbf{u}$ be the ion bulk flow velocity based on the thermal ions and $\delta \mathbf{v} = \mathbf{v} - \mathbf{u}$. Then
\begin{align}
\sum_s^{i,e}
\dxlb \frac{m_s}{2}
\int (\delta v^2 + u^2 + 2 \mathbf{u} \cdot\delta\mathbf{v})(\delta v_x + u_{x})f_s
d^3v \rb = 0,
\end{align}
The bulk velocity is approximately equal to the bulk flow velocity of only the thermal ions (an assumption that is supported by the simulations and by the fact that the number density of non-thermal particles is much less than the number density of the thermal particles). Expanding the polynomials gives
\begin{align}
\dxlb 
\sum_s^{i,e}
\frac{m_s}{2} 
\int (\delta v_x\delta v^2 + u^2\delta v_x + 2(\mathbf{u}\cdot \delta \mathbf{v}) (\delta \mathbf{v} \cdot \hat{x}) + u_{x} \delta v^2 + u_{x}u^2 + 2u_{x}\mathbf{u} \cdot \delta \mathbf{v})f_s d^3v \label{eq:energy_step}
\rb = 0,\\
\dxlb  Q_x + u_j P_{j,x} + \frac{1}{2}u_{x}\left ( P_{j,j} + \rho u^2\right )
\rb = 0,
\end{align}
where $P_{j,k} = \sum_s m_s \int (v_j - u_j)(v_k - u_k)f_s d^3v$ is the total pressure tensor (ion + non-thermals + electrons) and $Q_j = (m_i/2)\int (v_j - u_j)|\mathbf{v} - \mathbf{u}|^2f_i d^3v$ is the ion heat flux density (of the combined thermal and non-thermal ion distribution). 
We use the Einstein convention to imply summation over repeated indices.
Note that we take the limit where the electrons have negligible mass and are isotropic in the flow frame, which means that they only contribute to the total pressure.

We now study the effect of the energetic particles by decomposing the pressure into a thermal pressure $P_g$ and an energetic particle pressure $P_{nt,j,k}$ defined as
\begin{align}
P_{g} = \frac{m_i}{3}\int \delta v^2 f_{g} d^3 v + P_e,\\
P_{nt,j,k} = m_i\int (v_j - u_{nt,j})(v_k - u_{nt,k})f_{nt} d^3 v,
\end{align}
where $f_i = f_g + f_{nt}$ and $f_g$ and $f_{nt}$ correspond to the inflowing thermal and energetic non-thermal ion populations, respectively (we use the gas and CR notation for consistency with the astrophysical literature), $u_{nt,j}$ is the bulk velocity of the non-thermal particles, and $P_e$ is the electron pressure.
Note we assume that the thermal distribution is isotropic in the bulk flow frame.
Using these additional definitions and assumptions Eq.~\ref{eq:energy_step} becomes
\begin{align} 
P_{j,k} = \delta_{j,k} P_g + P_{nt,j,k},
\end{align}
where $\delta_{j,k}$ is the Kronecker delta and
\begin{align} 
\dxlb Q_x  + u_x\left(\frac{\Gamma}{\Gamma - 1}P_g + \frac{1}{2}\rho u^2\right) + 
\frac{m_i}{2}
\int (u^2\delta v_x + 2(\mathbf{u}\cdot \delta \mathbf{v}) (\delta \mathbf{v} \cdot \hat{x}) + u_{x} \delta v^2 + u_{x}u^2 + 2u_{x}\mathbf{u} \cdot \delta \mathbf{v})f_{nt} d^3v
\rb = 0. \label{eq:energy_step_2}
\end{align}
It should be noted that while we break the pressure tensor into thermal and non-thermal contributions which is assumed to be isotropic in each of their respective frames, the heat flux corresponds to the composite value based on both populations.
Now we make the substitution $\mathbf{w} = \mathbf{u}_{nt} - \mathbf{u}$ and $\delta \mathbf{v}_{nt} = \mathbf{v} - \mathbf{u}_{nt}$ and we have $\delta \mathbf{v} = \delta \mathbf{v}_{nt} + \mathbf{w}$. Using this we rewrite the integral as
\begin{align} 
=\int \{u^2(\delta v_{nt,x} + w_x) + 2[\mathbf{u}\cdot (\delta \mathbf{v}_{nt} + \mathbf{w})](\delta v_{nt,x} + w_x) \\
+ u_{x} (\delta v_{nt}^2 + \epsilon^2 + 2 \delta \mathbf{v}_{nt} \cdot \mathbf{w}) + u_{x}u^2 + 2u_{x}\mathbf{u} \cdot (\delta \mathbf{v}_{nt} + \mathbf{w})\}f_{nt} d^3v. \nonumber
\end{align}
Terms linear in $\delta v_{nt}$ vanish under integration which leaves
\begin{align}
\int [u^2w_x + 
2(\mathbf{u}\cdot \delta \mathbf{v}_{nt} \delta v_{nt,x} + \mathbf{u}\cdot \mathbf{w}w_x)
+ u_{x} (\delta v_{nt}^2 + \epsilon^2) + u_{x}u^2 + 2u_{x}\mathbf{u} \cdot \mathbf{w}]f_{nt} d^3v,
\end{align}
which, upon including the factor of $m_i/2$ from Eq.~\ref{eq:energy_step_2}, finally becomes
\begin{align}
= \frac{1}{2}\rho_{nt}u^2w_x + 
u_jP_{nt,jx} + 
\rho_{nt}\mathbf{u}\cdot \mathbf{w}w_x + 
\frac{1}{2}u_{x} P_{nt,j,j} + 
\frac{1}{2}\rho_{nt}u_{x}\epsilon^2 + 
\frac{1}{2}\rho_{nt}u_{x}u^2 +
\rho_{nt}u_{x}\mathbf{u} \cdot \mathbf{w}\\
= u_j P_{nt,jx} +
\frac{1}{2}u_{x} P_{nt,j,j} + 
\rho_{nt}\left(u^2w_x +
w_x\mathbf{u}\cdot \mathbf{w} +
\frac{1}{2}u_{x}\epsilon^2 +
\frac{1}{2}u_{x}u^2 +
u_{x}\mathbf{u} \cdot \mathbf{w}\right).
\end{align}
Normalizing this equation by $\rho_1 u_1^3$ lets us use the fact that each term multiplied by $\rho_{nt}/\rho_1$ is small compared to the non-thermal pressure contribution, so it can be neglected which simplifies the analysis considerably. Reintroducing this term back into Eq.~\ref{eq:energy_step_2} gives
\begin{align}
\dxlb Q_x  + u_x\left(\frac{\Gamma}{\Gamma - 1}P_g + \frac{1}{2}\rho u^2\right) +  \frac{1}{2}u_{x}P_{nt,j,j} + u_{j}P_{nt,j,x}
\rb = 0.\label{eq:apdx:jump}
\end{align}
The terms related to $P_{nt}$ represent the extra enthalpy flux of the non-thermal particles, being brought in towards the shock with the upstream flow.
Integrating this equation over a small region around the shock front 
yields Eq.~\ref{eq:energy}.

From panel (d) of Figure~\ref{fig:avg_jumps} it is clear that the heat flux density and the non-thermal enthalpy flux density terms cancel each other out. 
It can be shown that this result implies that the non-thermal particles must be at rest in the shock frame.
This result is consistent with what is found in Figure~\ref{fig:M}b, which shows the non-thermal distribution roughly centered around the origin in the shock frame near the shock front.
Furthermore, the condition that $u_{nt,x} = 0$ can also be connected to the steady-state solution to the Parker transport equation in the absence of non-thermal sources or sinks \citep[e.g.,][]{drury+81a,caprioli+09a}. From the non-thermal transport picture, this condition is associated with the advection flux of the non-thermal particles balancing the diffusive flux. 
The results presented in this manuscript are likely consistent with the established CR-modified shock phenomenology, without the need to invoke a ``sub-shock''.
In a forthcoming study we plan to re-derive the results in this manuscript in the fully relativistic limit and connect it to the standard Cosmic-Ray-modified shock framework.

Finally, this work assumes several times that the pressure in non-thermal particles is on the order of 10\% for the upstream kinetic ram pressure and that the non-thermal pressure remains roughly constant across the shock. While many previous works have used this assumption and demonstrated this in observations and simulations, we also include a plot of the averaged non-thermal pressure here. The assumptions and values are supported by what is found in the simulation, and we find a average value of the normalized non-thermal pressure of $\xi = 0.08$.

Eq.~\ref{eq:apdx:jump} should be valid for splitting the distribution function into any two populations, so long as the `nt' population has a sufficiently small number density. 
When combined with the assumptions used to derive Eqs.~\ref{eq:pflux} and \ref{eq:Eflux}, the solution will have some sensitivity to the specific energy cutoff that separates the two populations.
However, since the enthalpy and heat flux densities of the total ion distribution cancel in the upstream, the cutoff physically corresponds to the segment of the ion population that can reenter the upstream region from the downstream.
The nominal value of $10E_{sh}$ was found in previous works to correspond to just that \citep {caprioli+14a,caprioli+15}, and the strong agreement between our simulations and prediction based on this energy cutoff further supports this choice.
\begin{figure}
\centering
\includegraphics[width=0.65\textwidth,trim={0 0 0 0},clip]{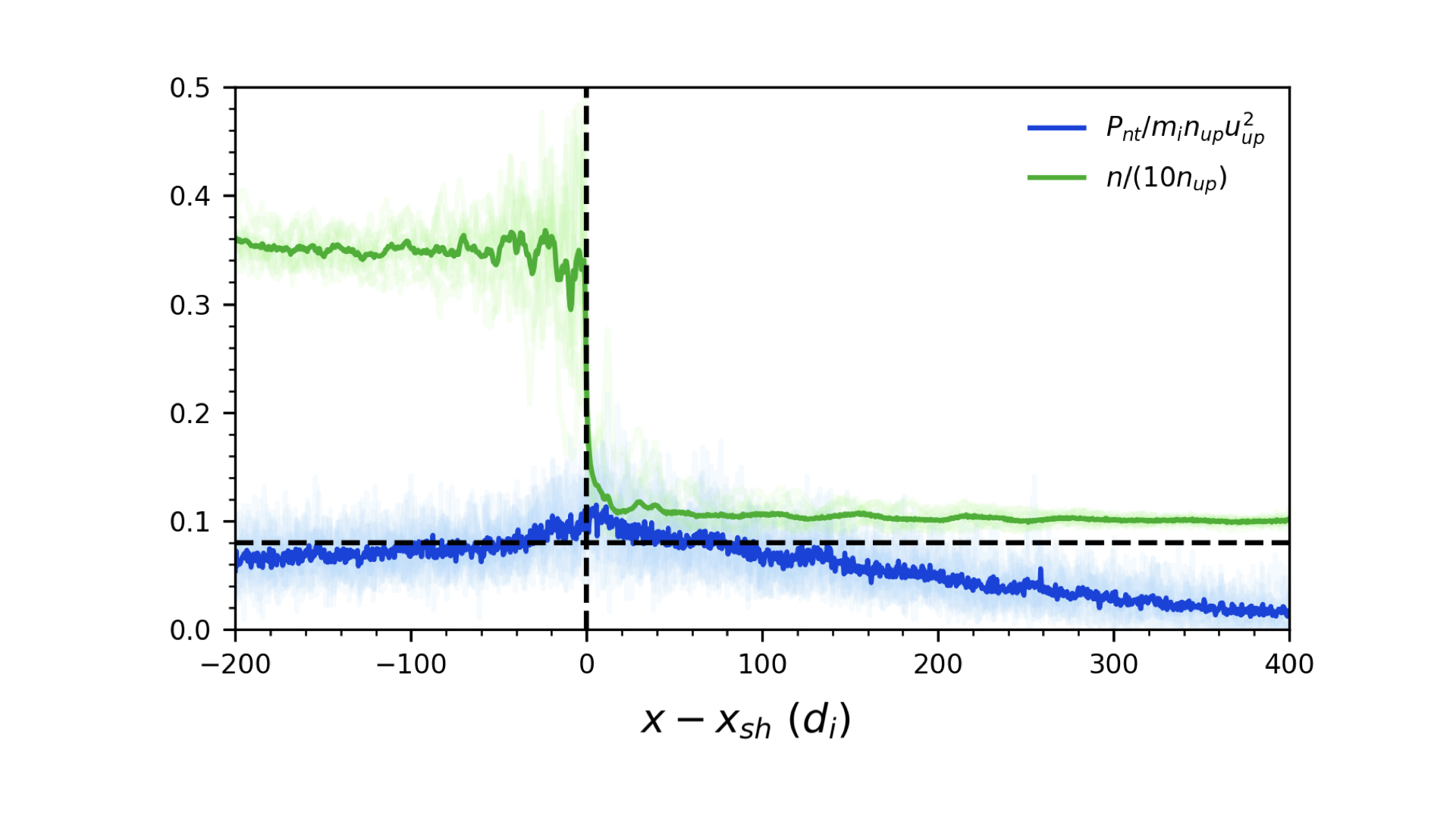}
\caption{
The non-thermal pressure and the total ion number density (divided by 10).
1D cuts are normal to the shock and centered around the shock front and are time-averaged as discussed in Figure~\ref{fig:Pnt}.
}
\label{fig:Pnt}
\end{figure}

\section{Jump Conditions with Balanced Heat Flux}
\label{apdx:jumpheatbal}
We now use the balancing of the upstream heat flux and non-thermal enthalpy flux to derive an updated set of jump conditions and predict the associated hydrodynamic modifications.
We start by considering the jump conditions for a generic distribution function which are given by
\begin{align}
\lb B_x \rbud = 0,\\
\lb \rho u_x \rbud = 0,\\
\lb  u_xB_y - u_yB_x \rbud  = 0,\label{eq:dHT}\\
\lb \rho u_x^2 + P_{x,x} + \frac{B_y^2}{8\pi} \rbud = 0, \label{eq:rhmom} \\
\lb \rho u_x u_y - \frac{B_xB_y}{4\pi} \rbud = 0,\\
\lb Q_x  + \frac{1}{2}u_{x} P_{j,j} + u_{j} P_{j,x} + \frac{1}{2}\rho u^ 2u_x +
\frac{B_y(u_xB_y - u_yB_x)}{4\pi} \rbud = 0. \label{eq:rhnrg}
\end{align}
Without loss of generality we take the x-direction as the shock normal direction and assume that the magnetic field lies in the $x-y$ plane.
We elect to perform the analysis in the de Hoffmann-Teller frame, which has the advantage of making all flow parallel to the upstream and downstream magnetic field (i.e., $\mathbf{u}\times\mathbf{B} = 0$).
This simplifies the set of equations considerably by eliminating Eq.~\ref{eq:dHT} and removing the magnetic field term from the energy equation.
The pressure term can be approximated as the sum of the thermal and non-thermal pressures, as discussed in the previous section, so Eqs.~\ref{eq:rhmom} and \ref{eq:rhnrg} become
\begin{align}
\lb \rho u^2 + P_g + P_{nt,x,x} + \frac{B_y^2}{8\pi} \rbud = 0,\\
\lb Q_x  + u_x\frac{\Gamma}{\Gamma - 1}P_g + \frac{1}{2}u_{x} P_{nt,j,j} + u_{j} P_{nt,j,x} + \frac{1}{2}\rho u^ 2u_x  \rbud = 0.
\end{align}
These jump conditions can be written out explicitly for the regions just upstream (1) and downstream (2) of the shock.
Note that subscript $1$ is for the upstream region corresponding to the ion foreshock (or non-thermal precursor) that extends $\sim 100 d_i$ upstream of the shock in this simulation.
Furthermore, we are assuming that the density and bulk flow in this region are approximately the same as that of the far upstream plasmas (where non-thermal particles have not yet reached).
This assumption necessarily breaks our assumption of shock stationary, however both conditions are found to be approximately satisfied in the simulation.
The downstream and foreshock region (i.e., region ``1'') are found to satisfy stationarity, while the far upstream violates it.
This is because a small fraction of non-thermal particles are escaping the shock and are continuously expanding into the far upstream region.

As discussed in Sec.~\ref{apdx:QPflux} the heat flux density and non-thermal enthalpy flux density in the upstream region cancels, and we take the downstream non-thermal distribution to be isotropic in the downstream frame, which yields
\begin{align}
\rho_1 u_{x,1} &= \rho_2 u_{x,2}\label{eq:Amflux}\\
\rho_1 u_{x,1}^2 + P_{g,1} + \frac{B_{y,1}^2}{8\pi} &= 
\rho_2 u_{x,2}^2 + P_{g,2} + \frac{B_{y,2}^2}{8\pi},\label{eq:Apflux}\\
\rho_1 u_{x,1}u_{y,1} - \frac{B_xB_{y,1}}{4\pi} &= 
\rho_2 u_{x,2}u_{y,2} - \frac{B_xB_{y,2}}{4\pi},\label{eq:Abflux}\\
u_{x,1} \left ( \frac{1}{2}\rho_1 u_1^2 + \frac{\Gamma}{\Gamma - 1}P_{g,1} \right ) &= 
u_{x,2} \left ( \frac{1}{2}\rho_2 u_2^2 + \frac{\Gamma}{\Gamma - 1}(P_{g,2} + P_{nt,2}) \right ).\label{eq:AEflux}
\end{align}
where we use $B_{x,1} = B_{x,2} = B_x$, and that the downstream non-thermal pressure is isotropic and non-relativistic.
The non-thermal pressure is expected to be roughly constant across the shock and so $P_{nt}$ does not appear in the momentum equation; this is because sufficiently energetic non-thermal particles are able to move across the shock freely and the non-thermal pressure is roughly continuous and only gradually varying (as is demonstrated in Figure~\ref{fig:Pnt}.
Any additional non-thermal contribution to the downstream heat flux density is neglected because the distribution is roughly isotropic.
Now, Eqs.~\ref{eq:Amflux},~\ref{eq:Apflux},~\ref{eq:Abflux}~and~\ref{eq:AEflux}, along with the condition that $\mathbf{u} \times \mathbf{B} = 0$, can be used to determine the downstream conditions for a given set of upstream parameters and a predetermined non-thermal pressure.
We do this following the standard approach by substituting in the following definitions
$r  \equiv \rho_2/\rho_x = u_x/u_2$, 
$M \equiv u_x\sqrt{\rho_x/\Gamma P_{g,x}}$,  
$\xi \equiv P_{nt,2}/\rho_x u_x^2$ and $R = P_{g,2}/\rho_xu_x^2$, and solving for $r$.
Through straightforward algebra the equations can be reworked into a quartic polynomial:
\begin{equation}
    c_4 r^4 + c_3 r^3 + c_2 r^2 + c_1 r + c_0 = 0,\label{eq:r:apdx}
\end{equation}
where
\begin{align*}
    c_4 &= -\frac{1}{\ma^4}\left[ \frac{1}{\gm\ms^2} + \frac{1}{2\cc}\right]\\
    c_3 &= \etag \frac{1}{\ma^4} \left[ \xi + \frac{1}{\Gamma \ms^2} + 1 + \frac{\tn}{2 \ma^2} \right] + \frac{1}{\ma^2} \left[ 
    \frac{2}{\gm\ms^2} + \frac{1}{\cc} - \etag \frac{\tn}{2}\left(1 - \frac{1}{\ma^2} 
    \right)^2 \right]\\
    c_2 &= \frac{-1}{\gm\ms^2} - \frac{1}{2\cc} 
    - \etag\frac{2}{\ma^2}\left[\xi + 1 +\frac{1}{\Gamma\ms^2} + \frac{\tn}{2\ma^2} \right] 
    + \frac{1}{\ma^4}\left(\frac{1}{2} - \etag\right) 
    +\frac{\tn}{2}\left[1 - \frac{1}{\ma^2}\right]^2 \\
    c_1 &= \etag\left[\xi + 1 + \frac{1}{\Gamma \ms^2} + \frac{\tn}{2\ma^2} \right]
    - \frac{2}{\ma^2}\left[\frac{1}{2} - \etag \right]\\
    c_0 &= \frac{1}{2} - \etag
\end{align*}
where $\ma = u_{x,1}/(B_x/\sqrt{4\pi\rho_1}) = (B_1/B_x)M_A = M_A/\cos{(\theta_{Bn})}$. This polynomial yields four roots, two of which are complex and only one of the remaining terms corresponds to an increase in the compression ratio.

In the parallel shock limit ($\theta_{Bn}=0$) this result can be shown to reduce to a much simpler form:
\begin{equation}
r = 
\frac{ \left( \Gamma + \Gamma \xi + \sqrt{1 + \psi\xi}\right)M^2 - 
(\sqrt{1 + \psi\xi}  - 1)}
{\left(\Gamma - 1\right)M^2 + 2}\label{eq:r:apdx_par},
\end{equation}
where
\begin{equation}
\psi = \frac{\Gamma^2 \left(2 + \xi + \frac{2}{\Gamma M^2}\right)}{(1 - \frac{1}{M^2})^2}
\end{equation}
and $M = M_s$.

In the limit of realistic efficiencies ($\xi \lesssim 0.1$), the square root term can be expanded and $\psi \approx 2 \Gamma^2$ yielding
\begin{equation}
    r \approx
    \frac{(\Gamma + 1)M^2}{2 + (\Gamma - 1)M^2} + 
    \xi M^2 \frac{\Gamma + \Gamma^2(1 + \frac{1}{M^2})}{2 + (\Gamma - 1)M^2},
\end{equation}
where the left term is the normal RH prediction with a maximum value of 4 and the right is the non-thermal pressure contribution.
The non-thermal contribution can be simplified further by noting that all of the terms without a leading factor of $M^2$ can be neglected as they are multiplied by the smaller term $\xi$, which gives
\begin{equation}
    r \approx
    \frac{(\Gamma + 1)M^2}{2 + (\Gamma - 1)M^2} + 
    \frac{\Gamma (\Gamma + 1)}{\Gamma - 1}\xi
\end{equation}
In the limit where the adiabatic index for the the non-thermal population is different than the gas, e.g., if the non-thermal distribution is relativistic, we expect the $\Gamma$ on the right most term to correspond to this value:
\begin{equation}
    r \approx
    \frac{(\Gamma + 1)M^2}{2 + (\Gamma - 1)M^2} + 
    \frac{\Gamma_{nt} (\Gamma_{nt} + 1)}{\Gamma_{nt} - 1}\xi
\end{equation}
In the absence of energetic particles ($\xi \rightarrow 0$) the expression simplifies to the standard hydrodynamic prediction.
Additionally, in the limit where $M$ is sufficiently large so that $1/M^2$ terms can be neglected, and using $\Gamma_{nt} = 5/3$, we find
\begin{align}
r \approx  4\left(1 + \frac{5}{3}\xi\right).
\end{align}


\end{document}